\documentclass[%
mathleft,%
]{an} 
\usepackage{graphicx} 
\usepackage{times} 
\begin{document} 
 
\Pagespan{1}{}
\Yearpublication{2012}%
\Yearsubmission{2012}%
\Month{1}%
\Volume{999}%
\Issue{92}%

\title{Solar irradiance variability} 
 
\author{Sami K.\ Solanki\inst{1,2}\fnmsep\thanks{Corresponding author: 
  \email{solanki@mps.mpg.de}} 
\and Yvonne C.\ Unruh\inst{3}} 
 
\titlerunning{Solar Irradiance Variability} 
\authorrunning{S.K.\ Solanki and Y.C.\ Unruh} 
\institute{Max Planck Institute for Solar System Research, Max Planck Str. 2, 
D-37191 Katlenburg-Lindau, Germany
\and 
School of Space Research, Kyung Hee University, Yongin, Gyeonggi 446-701, Korea 
\and 
Astrophysics Group, Blackett Laboratory, Imperial College, London, SW7 2AZ, 
United Kingdom 
}

\received{XXXX} 
\accepted{XXXX}
\publonline{XXXX}

\abstract{%
The Sun has long been considered a constant star, to the extent that its total
irradiance was termed the solar constant. It required radiometers in space to
detect the small variations in solar irradiance on timescales of the solar
rotation and the solar cycle. A part of the difficulty is that there
are no other constant natural daytime sources to which the Sun's brightness can
be compared. The discovery of solar irradiance variability rekindled a
long-running discussion on how strongly the Sun affects our climate. A
non-negligible influence is suggested by correlation studies between solar
variability and climate indicators.
The mechanism for solar irradiance variations that fits the observations 
best is that magnetic features at the solar surface, i.e. sunspots, faculae 
and the magnetic network, are responsible for almost all variations 
(although on short timescales convection and p-mode oscillations also 
contribute). In spite of significant progress important questions are 
still open. Thus there is a debate on how strongly irradiance varies on
timescales of centuries (i.e. how much darker the Sun was during the Maunder
minimum than it is today). It is also not clear how the solar spectrum changes
over the solar cycle. Both these questions are of fundamental importance for
working out just how strongly the Sun influences our climate.
Another interesting question is how solar irradiance variability compares 
with that of other cool dwarfs, particularly now that observations from 
space are available also for stars.
}

\keywords{Sun: magnetic fields, Sun: activity, 
Stars: activity, Stars: magnetic fields} 
 
\maketitle 

\section{Introduction} 

How strongly the Sun varies is of interest not just for solar physicists, but
also for astronomers, for whom the Sun has for many years been the
prototype of a constant star; its radiative output, or total solar
irradiance, has in the past generally been referred to as the solar constant.
This interest has been rekindled with the advent of high precision radiative
flux time series of large numbers of stars by the Kepler (Borucki et al.\,2010)
\nocite{borucki+2010} and CoRoT (Baglin et al.\,2006) \nocite{baglin2006} 
spacecrafts.

However, the strongest and most important reason for interest in solar variability 
is the Sun's potential influence on climate. As the only serious source of 
energy from outside Earth, the Sun provides basically all the energy 
needed to keep the Earth at its present temperature and hence is mandatory 
for maintaining higher life on Earth. 
 
For us denizens of the Earth, the main quantity of interest from the Sun is 
its irradiance, i.e. the radiative flux of the Sun measured at 1AU (in space, 
i.e. without the hindrance of the Earth's atmosphere). The irradiance is a 
spectral quantity and is often called the spectral irradiance. The spectral 
irradiance is not that easy to measure reliably, specially over longer times 
because spectrometers (and in particular their detectors) tend to degrade in 
sensitivity with time. More reliable to measure is the total solar irradiance 
(TSI), which is the integral of the 
spectral irradiance over all wavelengths, i.e. the total radiative energy 
flux from the Sun. Since almost all the energy emitted by the Sun is in the 
form of radiation, this is basically the total solar energy flux. Averaged 
over a solar rotation period, the TSI is roughly proportional to the solar 
luminosity, if we assume that the radiative output seen from the direction of 
the poles either remains constant or changes in phase with that seen from lower 
latitude vantage points. 
 
The Sun has been found to be variable on all timescales that we have so far 
been able to resolve or to cover. In the following we discuss such 
variability divided between different timescales and consider also briefly 
similar variability on Sun-like stars. 
 
\section{Overview of solar irradiance variability and its causes} 
 
\begin{figure}
\includegraphics[width=\linewidth,height=53mm]{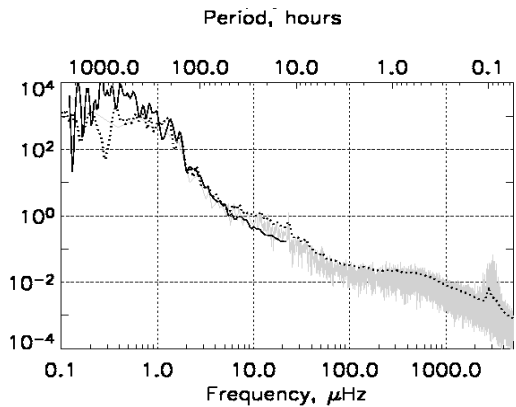}
\caption{Power spectrum of solar total irradiance showing the Fourier 
(grey line) and global wavelet (black dotted line) power spectra 
(in ppm$^2 /60 \mu$Hz) of the VIRGO data set for the year 2002, sampled 
at a 1-min cadence \nocite{frohlich+1995} (Fr\"ohlich et al.\,1995). 
The black solid line shows the global wavelet spectrum of the SORCE 
TIM data \nocite{kopp+2005} (Kopp, Lawrence \& Rottman 2005) for 
the year 2003 sampled every 6 h \nocite{seleznyov+2011}
(from Seleznyov et al.\,2011).}
\label{fig:power}
\end{figure}

To begin, let us briefly discuss the sources of variability at 
different timescales. Figure~\ref{fig:power} 
shows the power spectrum of total solar 
irradiance for periods from around 1 minute to roughly 1 year. Clearly, power 
is larger at longer periods and drops rapidly towards shorter timescales. 
The sources of the power are also quite varied, ranging from p-modes, which 
give the peak at around 5 min (3000 $\mu$Hz),  granulation (leading to the 
plateau between 50 and 500 $\mu$Hz), as well as rotational modulation 
due to the passage of sunspots and faculae/plages over the solar 
disk (the rotation frequency of the Sun lies at 0.45 $\mu$Hz). No significant 
peak is 
seen at the rotation period itself, because most sunspots live less long than 
a full rotation and larger active regions (which do live longer) evolve in this
time period and are rarely present on their own (the steep 
rise between 1 and 10 $\mu$Hz is due to the combined effect of rotational and 
evolutional effects). 
 
Thus irradiance variability on timescales of minutes to hours is driven by 
granulation and $p$-mode oscillations, from hours to days by the evolution 
and rotational modulation of magnetic features (specifically sunspots), 
with a contribution by granulation (and possibly supergranulation). Between
days and the solar rotation period the main contributors are sunspots, whose
darkness is rotationally modulated (but also faculae). From the solar
rotation period to the solar cycle the variability is determined mainly by
active region evolution, the appearance and disappearance of active regions
and their remnants, and finally the solar cycle itself. On centuries to
millennia the long-term evolution of the magnetic field, produced by the
solar dynamo, displays secular changes (including marked changes on time
scales of 70-100 years - the so-called Gleissberg cycle) that are 
expected to give rise to secular variations in the irradiance. On still 
longer timescales up to $10^5$ years, the thermal relaxation 
timescale of the convection zone, the energy blocked by sunspots is 
released again, while beyond $10^6$ years solar evolution leads to a 
slow brightening of the Sun.

Most of the causes of irradiance variations listed above are associated more 
or less directly with the Sun's magnetic field (with the exception of the 
very shortest and the very longest timescales), mainly via the influence of 
sufficiently strong magnetic fields on the thermal structure of the solar 
surface and atmosphere. 
 
Magnetic concentrations on the solar surface can be dark 
(e.g. large magnetic features, sunspots, pores) or bright (the small 
magnetic features forming faculae and the network). Since magnetic 
flux concentrations are well described by flux tubes (Yelles Chaouche, 
Solanki \& Sch\"ussler 2009), \nocite{yelles+2009} the question arises why 
some flux tubes are bright while others are dark.

Two counteracting processes determine the surface brightness of solar 
magnetic features. The strong magnetic field within both small and large 
magnetic flux tubes reduces the convective energy flux, 
which leads to a cooling of the magnetic features, since the 
vertical radiative energy flux in the convection zone is comparatively 
small and cannot fully compensate the reduction in convective flux. 
 
The large magnetic pressure in the interiors of magnetic flux concentrations 
and horizontal pressure balance implies that magnetic features are evacuated. 
Consequently, the evacuated magnetic structures are heated 
by radiation flowing in from their dense and generally hot 
walls. This radiation  heats flux tubes narrower than 
about 250 km sufficiently to make them brighter than the mainly field-free 
part of the photosphere, especially when seen near the 
limb where the bright interfaces of the flux tubes with their surroundings 
are best visible \nocite{spruit1976,keller+2004} (Spruit~1976; Keller et 
al.~2004). For larger features the radiation flowing in from the
sides penetrates only a very small part of the volume of the magnetic feature (the 
horizontal photon mean free path is roughly 100 km), 
so that features greater than roughly 400 km in diameter remain 
\nocite{grossmann-doerth+1994} dark (pores and sunspots), cf.~Grossmann-Doerth 
et al.\,(1994). 
 
According to \nocite{spruit1982} Spruit (1982),  the energy blocked by sunspots gets 
redistributed in the solar convection zone and is re-emitted 
again slowly over the convection zone's Kelvin-Helmholtz timescale, 
i.e. around 100, 000 years. Similarly, the excess radiation coming 
from small flux tubes is also taken from the heat reservoir of the 
convection zone. In this sense these small, evacuated magnetic features act as 
leaks in the solar surface, by increasing the solar surface area. 

\section{Irradiance variations at timescales up to the solar cycle}
Reliably measured changes in the TSI were first reported in the
early 1980s 
and soon attributed to sunspots and faculae \nocite{willson1982}
(e.g., Willson~1982). TSI has been monitored continuously since 1978,
and a number of TSI composites that combine measurements
from a sequence of different instruments are available
(see, \nocite{domingo+2009} Domingo et al.\,2009 for a discussion of 
the different composites).
These composites agree well with each other on timescales
ranging from days to years, though the stability of the instruments
contributing to the composites is not yet good enough to
unambiguously detect changes (or indeed the absence of changes)
between subsequent cycle minima.

\begin{figure}
\includegraphics[height=\linewidth,angle=90]{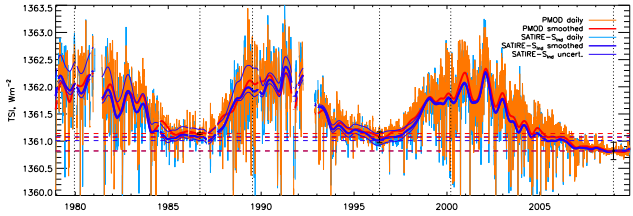}
\caption{Modelled and observed total solar irradiance from 1978 to 2010. 
The thin orange and light-blue lines are daily values from the PMOD 
composite (Fr\"{o}hlich 2000) and SATIRE-S (Ball et al.\,2012) 
\nocite{frohlich2000,ball+2012tsi}, respectively.
The thick red and blue lines show the smoothed PMOD and SATIRE-S values. 
Figure kindly provided by W.~Ball.
}
\label{fig:tsi}
\end{figure}

Just as for Sun-like stars, rotational variability on the Sun
is dominated by cool spots; these typically produce 
dimmings of the order of 1000 ppm that can be clearly seen on 
Fig.~\ref{fig:tsi} where TSI is plotted for the last three solar 
cycles. Note that some large spot groups can lead to more dramatic 
effects, such as the decrease of approximately 4000~ppm that 
was produced by the passage of the sunspot group that gave rise
to the 2003 `Halloween storm'.
In addition to the spot dimmings, the Sun also shows brightenings 
due to faculae; these brightenings generally lead to increases 
of about 200 ppm (e.g., Fligge, Solanki \& Unruh 2000) \nocite{fligge+2000} 
in the TSI from a typical active region. 

\begin{figure}
\includegraphics[width=\linewidth,height=43mm]{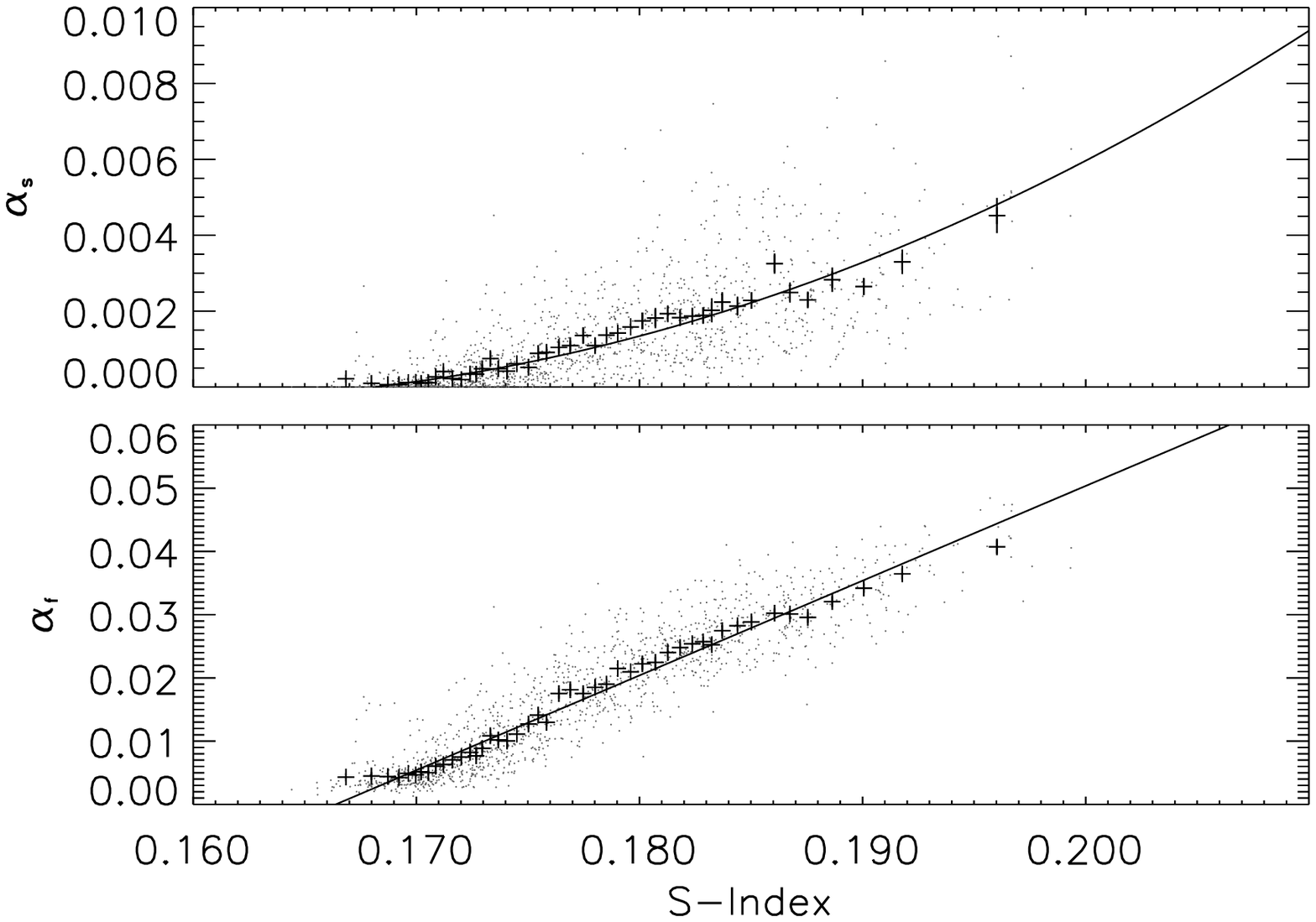}
\includegraphics[width=\linewidth,height=42mm]{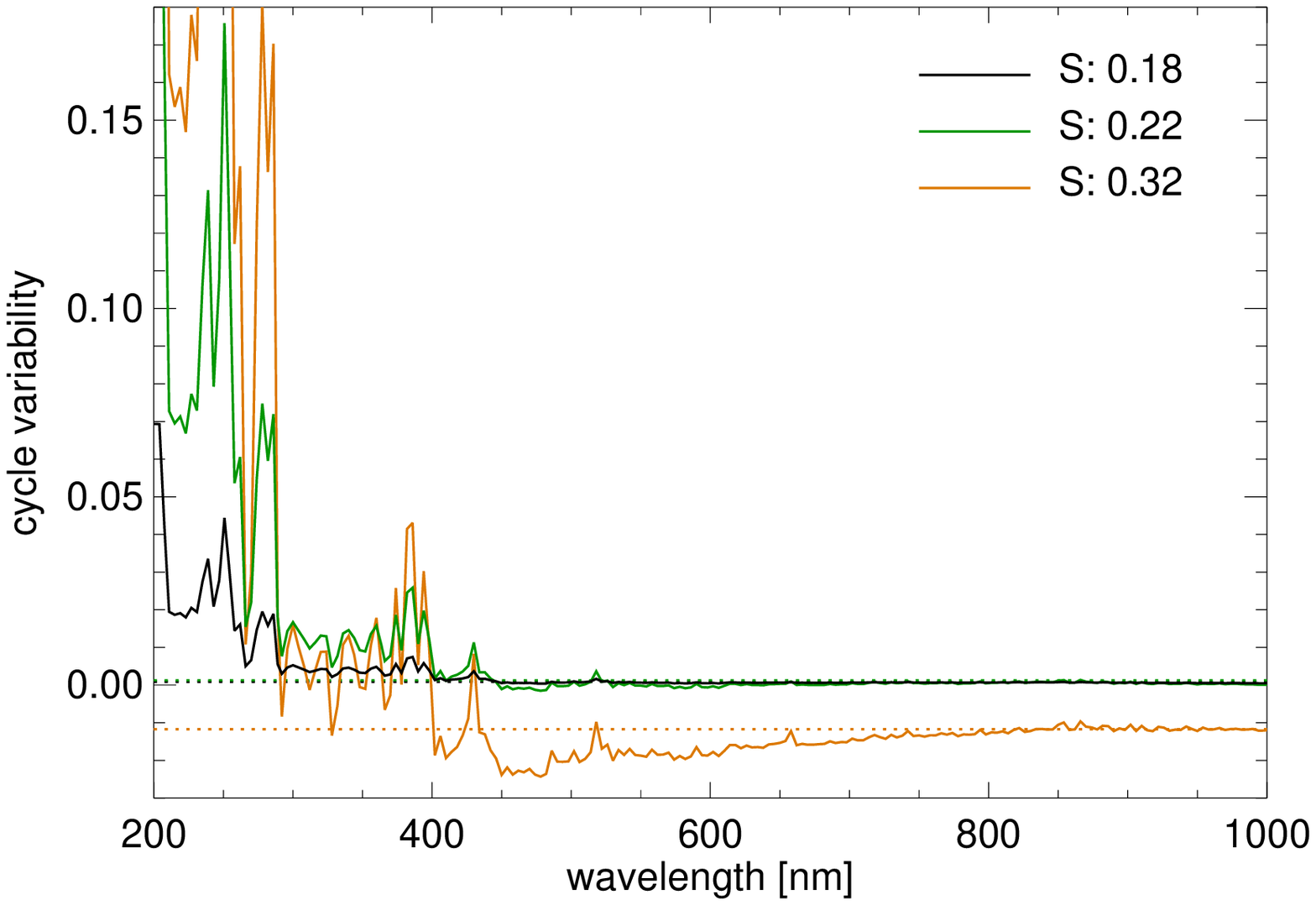}
\caption{{\em Top:} solar facular and spot filling factors as a function of
the Ca~S-index (filling factors are from Wenzler et al.\,2006; the S-index has
been computed from the Kitt Peak Ca data). Figure kindly provided R. Knaack.
{\em Bottom:} Cyclical variability as a
function of wavelength for hypothetical Sun-like stars with maximum
S-indices of 0.18 (black line), 0.22 (green line) and 0.32 (orange line). 
The S-index at minimum is 0.167 in all three cases. The dotted lines 
indicate the
wavelength-integrated variability (on this scale the total variability 
for S-indices of 0.18 and 0.22 can not be distinguished). 
}
\label{fig:starvar}
\end{figure}

As the Sun goes from sunspot minimum to maximum, more
flux emerges on the solar surface, and the area coverage
of bright flux tubes initially increases more rapidly than that 
of the spots. This is illustrated in Fig.~\ref{fig:starvar} (top) where 
the sunspot and facular filling factors are plotted as a function 
of the solar S-index, a measure of the solar activity. Combined with the 
longer life times of the faculae, their larger area coverage results 
in the Sun being `activity-bright' on cyclical timescales, i.e., 
the Sun is (on average) brighter during sunspot maximum than during 
sunspot minimum. Sunspot darkening and facular brightening are in 
a delicate balance on the Sun; indeed, there are good indications that 
the Sun is darker during times of higher activity in some wavelength 
bands \nocite{unruh+2008} (Unruh et al.\,2008).

A number of empirical and semi-empirical irradiance models have been developed.
The empirical models generally are simple regressions between one or more
activity indicators and the TSI, while the semi-empirical models make use of
model atmospheres to compute the full solar spectrum and determine the TSI
from the surface distribution of magnetic (or brightness) features on
the solar surface. The models successfully reproduce TSI changes on
timescales ranging from days to years and show that more than 90~\% of
the TSI changes on timescales of weeks to years can be explained as being
due to changes in the solar surface flux
\nocite{krivova+2003_cycle23,wenzler+2006,preminger+2010,ball+2012tsi}
(Krivova et al.\,2003; Wenzler et al.\,2006; Preminger et al.\,2010; Ball
et al.\,2012). An example is shown in Fig.~\ref{fig:tsi} where the TSI
modelled with SATIRE-S is plotted alongside the PMOD composite
\nocite{frohlich2000} (Fr\"ohlich~2000).

Changes in the solar irradiance show a strong wavelength dependence 
and measuring solar spectral irradiance (SSI) changes accurately is
challenging. For wavelengths below 400~nm, a reasonably long data train
is available from a succession of instruments, the longest being from
the SU\-SIM and SOL\-STICE instruments onboard the Upper Atmosphere Research
Satellite, UARS (see Deland \& Cebula (2008) \nocite{deland+cebula2008}
for a compilation of UV irradiances up to 2005). With the exception of
measurements in three narrow wavelength bands, data above 400~nm
only became available with SCI\-A\-MA\-CHY onboard EN\-VI\-SAT 
\nocite{skupin+2005} (since Aug 2002, Skupin et al.\,2005) and 
SIM onboard SORCE \nocite{rottman+2005} (since 
Apr 2004, Rottman et al.\,2005).

\begin{figure}
\includegraphics[width=\linewidth]{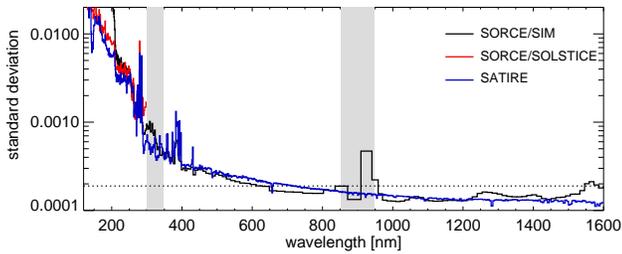}
\caption{Normalised standard deviation of the SSI for 3 solar rotations in
2005. The red line
shows the results from SORCE/SOLSTICE (v11 release), the black line those
from SORCE/SIM (v17 release). The blue line is the normalised standard
deviation obtained with SATIRE-S. The grey shading indicates the 
location of the SORCE/SIM detector edges.}
\label{fig:rotvar}
\end{figure}
On rotational timescales, there is good agreement between
the SSI changes measured with different instruments and models. We 
can quantify the spectral variability in terms of the normalised standard 
deviation (Johns \& Basri 1995) \nocite{johns+basri1995} which is 
given by $\sigma_\lambda / \bar{f_\lambda}$, where $\bar{f_\lambda}$
is the average irradiance at a given wavelength and
$\sigma_\lambda$ is the standard deviation, $[\sum_{j=1}^N (f_\lambda(t_j) - 
\bar{f_\lambda})^2/(N-1)]^{1/2}$. This 
quantity is plotted in Fig.~\ref{fig:rotvar} for three solar 
rotations in 2005 for the SATIRE-S model 
(blue line, Krivova et al.\,2003; Krivova, Solanki \& Wenzler 2009), for 
\nocite{krivova+2003_cycle23, krivova+2009tsi} 
SORCE/SOLSTICE (red line) and SORCE/SIM (black line). 
The variability generally decreases with increasing
wavelengths. The spectral variability is comparable to 
that of the TSI for wavelengths above 400~nm, but exceeds 
it by a factor of almost 10 below 240~nm, and can reach 
very high values in some individual lines, as, e.g., in the 
Mg~{\sc ii} line at 280~nm. The character of the variability 
also changes with wavelength. In the
UV below approximately 290~nm, variability is mostly due to faculae;
faculae and spots contribute roughly equally between 290 and 400~nm, while
spots dominate in the visible and NIR \nocite{unruh+2008} (Unruh et al.\,2008).

SSI changes on timescales of years and longer are less well 
determined because the uncertainty of the instrumental degradation 
corrections is of a similar order of magnitude as the expected 
long-term change at many wavelengths, particularly for wavelengths 
greater than 300~nm. Measurements taken with SORCE/SIM during the 
declining phase of cycle 23 
suggest that the decline in the UV irradiance (integrated over a 
wavelength range of 200 to 400~nm) exceeds the decline in the TSI, 
and that this excess decline is compensated by an anti-cyclic 
increases in the irradiance for wavelengths between 400 and 800~nm 
and 1000 and 1600~nm. This scenario is at odds with modelling results 
(e.g., \nocite{ball+2011,lean+deland2012} Ball et al.\,2011; Lean \& DeLand
2012)  and does not agree with previous measurements such as UARS/SUSIM 
where Morrill, Floyd \& McMullin (2011) \nocite{morrill+2011susim} estimate 
the UV contribution to the TSI to be approximately 60\% (see 
\nocite{ermolli+2012} Ermolli et al.\,2012 for further discussion).

\section{Irradiance variations on longer timescales: secular change}
One of the most contentious questions in current solar irradiance research
deals with the magnitude of secular variations in solar irradiance on time
scales of centuries up to millennia. Do such variations exist and how
strong are they? This question is particularly important, because this
magnitude partly decides on the effect of the Sun on climate and whether the
Sun has contributed to global warming.

\begin{figure}
\includegraphics[width=\linewidth,height=32mm]{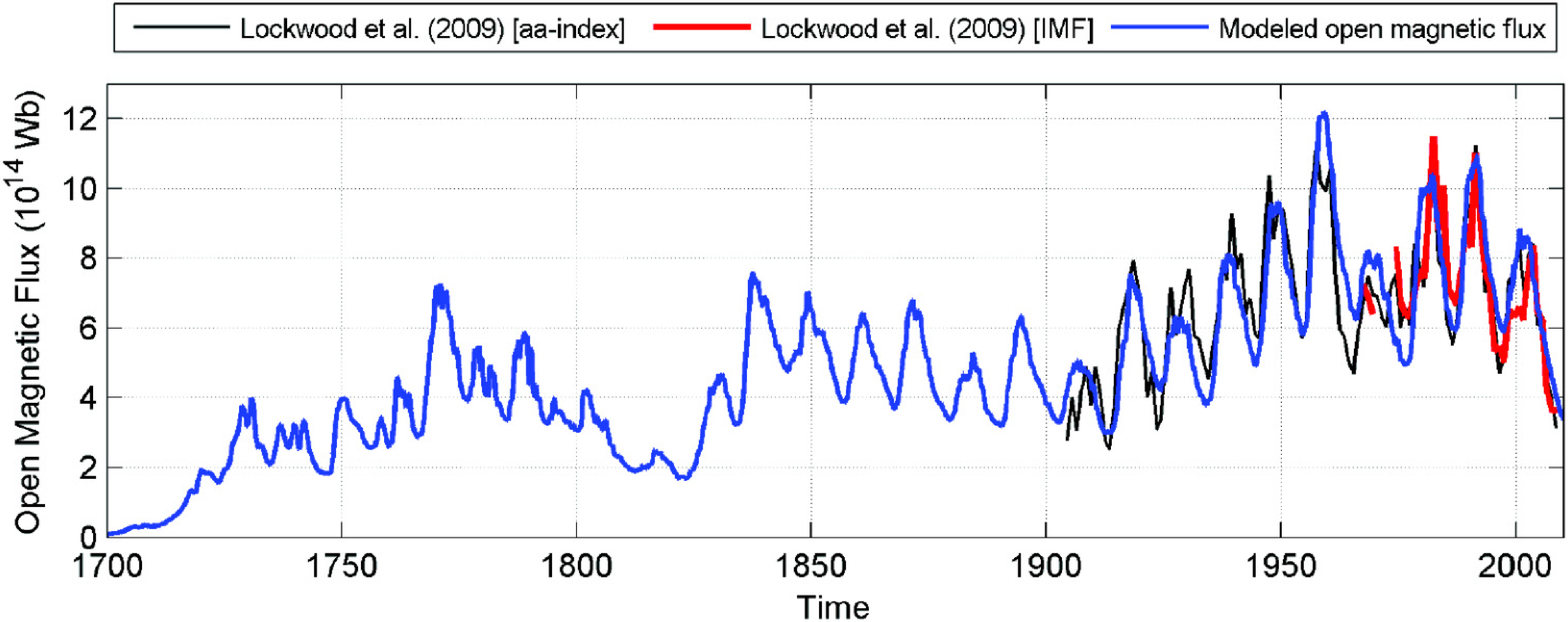}
\caption{Evolution of the Sun's reconstructed and modelled open
magnetic flux since the end of the Maunder minimum. The flux reconstructed
from the measured interplanetary magnetic
field is plotted in red, the reconstruction of the open flux from the
geomagnetic aa-index is plotted black (Lockwood et al.\,2009a,b) while the
open flux computed by the SATIRE-T model by Vieira \& Solanki (2010) is in
blue (adapted from Vieira et al.\,2011).}
\label{fig:open_flux}
\end{figure}

The only direct evidence for secular variations in irradiance comes from the
TSI record and the slightly lower brightness during the just completed activity
minimum between cycles 23 and 24 (see Fig.~\ref{fig:tsi}). However, there
is also other, more indirect evidence. In Fig.~\ref{fig:open_flux} we 
display the open magnetic flux of the Sun since 1900 (measured values from 
Lockwood et al.\,2009a,b 
\nocite{lockwood+2009openflux1,lockwood+2009openflux2} 
in black, modelled values from Vieira \& Solanki 2010 
\nocite{vieira+solanki2010} in blue). In
addition to the clearly visible cyclic changes, an increase in the first half
of the record is seen: the Sun's open magnetic flux roughly doubled between 
1900 and the middle of the 20th century. Then, during the deep and long 
minimum between cycles 23 and 24, the open flux dropped back to the level it 
had over a century ago. This is a large and rapid change. However, it is 
not clear to what extent and how strongly such variations in open flux 
convert into variations of the irradiance. Remarkably the model reproduces 
the observed variations rather well.

Therefore one of the major uncertainties in our knowledge of solar irradiance
is how strongly the solar irradiance varied since the Maunder minimum, with
the extreme values being 0.7 W\,m$^{-2}$ \nocite{schrijver+2011}
(Schrijver et al.\,2011)
and 6 W\,m$^{-2}$ \nocite{shapiro+2011} (Shapiro et al.\,2011).
Most reconstructions and estimates lie between these extremes 
\nocite{krivova+2007} (e.g. Krivova, Balmaceda \& Solanki 2007), i.e. 
roughly between 1 and 2 W\,m$^{-2}$ \nocite{schnerr+spruit2011} (e.g. the 
upper limit set by Schnerr \& Spruit 2011, although they only considered 
continuum brightness of magnetic features in the visible and thus 
underestimate the total 
brightness excess due to magnetic fields in the quiet Sun).

Once the magnitude of the variations has been determined, the variation with 
time can be found from a proxy of 
solar activity available for a sufficiently long period of time. Two 
important directly measured proxies are the sunspot area, available since 1874 
(Balmaceda et al.\,2009) \nocite{balmaceda+2009} and sunspot number that is 
available since 1610 \nocite{hoyt+schatten1998} (Hoyt \& Schatten 1998). For 
many purposes it is important to know the evolution of solar activity over an 
even longer period of time, e.g. in order to study the long-term 
development of the solar dynamo (how often do grand minima such as the 
Maunder minimum take place?) and in order to have better statistics on 
whether solar activity influences climate. 
 
Past solar activity can be reconstructed from cosmogenic isotopes, such 
as $^{10}$Be (Beer et al.\,1992) \nocite{beer+1992} or $^{14}$C 
\nocite{solanki+2004} (Solanki et al.\,2004). The 
former can be gleaned from ice cores drilled in Greenland or
Antarctica, the latter is obtained from tree trunks 
\nocite{stuiver+1998,reimer+2004} (and dated via
dendrochronology; Stui\-ver et al.\,1998, Reimer et al.\,2004). Recently, 
evidence has been presented that also nitrate in certain antarctic ice 
cores carries information on past solar activity \nocite{traversi+2012}
(Traversi et al.\,2012).

Reconstructions of, e.g., open magnetic flux or sunspot number from such data
(e.g. Usoskin et al.\,2004; Steinhilber, Beer \& Fr\"ohlich 2009)
\nocite{usoskin+2004,steinhilber+2009} reveal that there have
been regular episodes of extremely low solar activity (grand minima) as well 
as equally common periods of particularly high solar activity (grand maxima), 
e.g. Usoskin et al.\,(2007)\nocite{usoskin+2007}.
Particularly striking is that during the last 
roughly 70 years the Sun has been in a  prominent grand maximum, 
although it is agreed that the Sun is now leaving this phase 
\nocite{lockwood+2009grandmin,solanki+krivova2011}
(Lockwood, Rouillard \& Finch 2009; Solanki \& Krivova 2011).

Once an appropriate solar activity proxy has been determined, it can be used
to compute the solar total irradiance \nocite{steinhilber+2009,vieira+2011}
(Steinhilber et al.\,2009; Vieira et al.\,2011).
Such irradiance data sets display ups and downs associated with grand 
minima and maxima. The exact temporal variation depends mainly on the data set 
underlying the solar activity reconstruction (and to a smaller extent on the 
model used), while the amplitude of the irradiance variations must be set 
independently. 

\section{Comparison with other stars} 
Historically, ground-based stellar variability studies tended to 
focus on the most active stars with clear rotational 
cycles and dominated by large spots, some of them covering 
as much as 50~\% of the visible stellar 
hemisphere (see, e.g., \nocite{berdyugina2005} Berdyugina~2005
for an overview). 
 
An important step in understanding stellar variability for 
less active stars was the establishment of the 
Mt~Wilson Ca~{\sc ii} H\&K survey 
\nocite{wilson1978CaHKsurvey,baliunas95mtwilson} (Wilson 1978; Baliunas 
et al.\,1995) that showed that
activity cycles are ubiquitous on Sun-like stars with typical 
cycle lengths of 2.5 to 20~years. The cores of the 
Ca~{\sc ii} H\&K are good tracers of the magnetic field and scale with the 
magnetic field strength. A further 
milestone was the start of a photometric program 
in 1984 that monitored a large fraction of the Mt~Wilson targets 
in the $b$ and $y$ filters \nocite{lockwood+1997,radick+1998,lockwood+2007}
(Lockwood et al.\,1997, 2007; Radick et al.\,1998).
This has produced almost two decades of simultaneous chromospheric 
and photospheric variability data, allowing the Sun to be placed 
in the stellar context. Note that the photometric program measures the 
radiation arising in the photosphere, in contrast to the chromospheric 
information carried by the Ca~{\sc ii} H\&K line cores. 

Considering first the short-term variability that is thought to be 
largely due to stellar rotation, Lockwood et al.\,(1997) and Radick et al.\,(1998)
showed that stars with larger chromospheric emission (as
measured by the mean $\log R'_{HK}$ or the S-index) were generally
more variable with larger photometric ($b+y$) amplitudes.
Almost all stars show Sun-like rotational variability in the
sense that chromospheric and photospheric brightness are
anticorrelated, pointing to spots as the main causes of short-term
variability.
 
Lockwood et al.\,(1997) found that variability was more likely
for later spectral types; approximately 40\% of the G and K-type
stars in their sample were variable. This dependence
was recently confirmed by Basri et al.\,(2011) \nocite{basri+2011} who
used the exquisite precision offered by Kepler to investigate
stellar variability. They found that approximately half of all
Sun-like stars showed rotational variability (the slightly higher
variability fraction is to be expected for the higher precision
of the space-based Kepler observations).

Collating two decades of 
photospheric and chromospheric measurements, 
Lockwood et al.\,(2007) were able to show that stars fall into 
two groups in terms of their cyclical behaviour. The most active 
stars (as traced by their Ca~{\sc ii}~H\&K emission) tend to be 
`activity dark', i.e., they become fainter as their magnetic 
activity ramps up, while less active stars tend to follow the solar 
`activity-bright' pattern. 

Considering the Sun's total irradiance, the Sun is activity bright 
on the solar-cycle timescale, though it is located not far from the 
activity dark/bright boundary. However, comparing the Sun to 
other Sun-like stars is not entirely straightforward, as we do not 
yet have long-term solar spectral measurements that are equivalent 
to the $b$ and $y$ filters. In addition, we monitor the Sun from a 
somewhat special `equator-on' position, while the stellar sample 
will present a mix of inclination angles 
\nocite{knaack+2001} (see Knaack et al.\,2001). 

To shed more light on the expected spectral behaviour of Sun-like stars,
we consider the case of hypothetical Suns by extrapolating the
observed facular and spot filling factors shown in the top two 
plots of Fig.~\ref{fig:starvar} to higher activity levels.
The fits to the binned data (solid lines) suggest that the total
area covered by faculae increases linearly with the S-index, while
the spot areas increase as the square of the S-index.
Taking the typical mean S-index during solar maximum to be approximately 
0.18, the solid line in the bottom plot of Fig.~\ref{fig:starvar}
shows the expected solar-cycle variability for a simple 3-component
model of faculae, sunspots and quiet-Sun. Note that the active-region
contrast used here are derived from disk-integrated fluxes and
thus neglect the location of the active regions. Nevertheless,
the resulting spectrum is in good agreement with the more detailed
SATIRE-S model discussed earlier and presented in, e.g., \nocite{ball+2011}
Ball et al.\,(2011). Note that in this model the Sun is clearly activity bright 
in the UV; in the visible, it shows very little variability, though 
it remains marginally activity bright (though see \nocite{preminger+2011}
Preminger et al.\,2011). In the NIR it reverses sign (see also 
\nocite{solanki+unruh1998} Solanki \& Unruh 1998) and 
becomes activity dark (not shown here). 

Extrapolating the facular and spot filling factors to more extreme
values, we find that the behaviour of the hypothetical Suns switches 
from activity bright to activity dark in the visible while remaining 
activity bright at most UV wavelengths. The wavelength-integrated 
variability is not necessarily a good tracer of the spectral variability:
regardless of the very different spectral shape shown by the black 
and green curve, the integrated change is of the order 1000~ppm for 
$S=0.18$ and $S=0.22$. At $S=0.32$ (shown in orange), the star 
becomes clearly activity dark with a total irradiance change of 
-1.2~\%. 
 
\section{Conclusions} 
 
Solar irradiance variability has been mainly studied under the aspect of the 
Sun's possible influence on the evolution of the Earth's climate. However, it 
is  also becoming important for comparison with and to help in the interpretation 
of the microvariability of other cool stars, that has become a lot more 
accessible through the CoRoT and Kepler missions. 
 
Solar total irradiance has been measured since 1978 and we now have a 
reasonable understanding of the causes of TSI variations on different time 
scales. In particular, on the prominent timescales of solar rotation and the 
solar activity cycle, it is now clear that the TSI variations are caused by 
magnetic fields at the solar surface. Regarding the solar spectral irradiance 
variations the situation remains less clear. Models and the measurements by 
the SIM instrument on the SORCE spacecraft diverge rather strongly, even 
showing opposite phases of evolution 
in the 
visible part of the spectrum. The cause of this discrepancy remains one of 
the major open questions in solar irradiance research. 
 
Thanks to the breakthroughs of the last decade in the reconstruction of solar 
activity, there 
have been multiple reconstructions of solar total irradiance over the 
Holocene. These differ from each other in various aspects, but most strongly 
in the amplitude of the TSI changes from the Maunder minimum to the present 
day, with extreme estimates of this value diverging by an order of magnitude 
from each other. This quantity is a key to deciding how large the 
influence of solar irradiance variations on climate change has been and is 
expected to be in the future. Consequently, the uncertainties in this quantity 
constitute the second major unknown in our knowledge of solar irradiance 
variations. 

Besides these two major open questions, the main missing ingredient from the 
solar side in Sun-climate research is our inability to predict the course of 
future irradiance variations, or solar activity as a whole for that matter. 

\acknowledgements 
We thank N.A.\,Krivova and W.T.\,Ball for helpful discussions.
This work has been partially supported by the WCU grant (No.\,R31-10016)
funded by the Korean Ministry of Education, Science and Technology.

\bibliographystyle{an}

\end{document}